\documentclass{article}
\usepackage{natbib}
\usepackage{graphics}
\usepackage{graphicx}
\usepackage{latexsym}
\usepackage{a4}
\usepackage{a4wide}
\begin{document}

\hspace*{10.0cm}\parbox{4.0cm}{
OCU-HEP 2003-01 \\
}

\begin{center}
\begin{Large}
\begin{bf}
Cosmic-Ray Detection System using Internet
\end{bf}
\end{Large}
\end{center}

% Author list

\begin{center}
T.~Hamaguchi$^{a}$, M.~Katsumata$^{a}$, E.~Nakano$^{a}$, 
T.~Takahashi$^{a}$, Y.~Teramoto$^{a}$, 
Y.~Saito$^{b}$, Y.~Sasaki$^{c}$, M.~Honda$^{d}$, Y.S.~Honda$^{d}$
\end{center}

\begin{center}
\begin{it}
$^{a}$Institute for Cosmic Ray Physics, Osaka City University,
Osaka 558-8585, Japan

$^{b}$Osaka Science Museum, Osaka 530-0005, Japan

$^{c}$Izumiootsu High School, Osaka 595-0012, Japan

$^{d}$Kinki University Technical College, Mie 519-4395, Japan
\end{it}
\end{center}

\begin{abstract}
A cosmic ray detection system, consisting of standardized 
detector stations connected through the Internet, is described. 
The system can be used for detecting air showers that arrive over a wide area
with correlated time. The data at each site are exchanged in
(quasi) real time, which makes the system suitable for displaying
cosmic ray arrivals at museums and schools.
\end{abstract}


\input{epsf}
\section{Introduction}
\label{sec:introduction}

Hints of a time correlation between cosmic ray air showers have been
observed using detectors placed over a $100 \times 100$ km$^{2}$ area 
\cite{carrel,kitamura}. Kitamura et al. \cite{kitamura}
reported evidence for time-correlated events that were detected by 
two air shower arrays located 115 km apart; 
one at Kinki University, located in Osaka, and 
the other at the Mitsuishi Observatory of Osaka City University, 
located in Okayama. 
This report raised the question of the possible existence of 
high energy cosmic ray phenomena that extend over large distances. 
In the United States and Canada, a network of cosmic
ray detector: the North American Large Area Time Coincidence
Array (NALTA) has been built and is now being 
operated \cite{physics-today,fermi-news}.
In Europe, before the large LEP detectors were closed, 
studies of cosmic ray muons had been done by the 
ALEPH and L3 detectors with an extended air shower array 
\cite{cern-courier,aleph-nim}. 
Our project was started with the similar motivation. 
The system consists of standardized detector stations. 
Each station has four scintillation counters and it can locally detect 
and reconstruct air showers. 
We call this station a ``cosmic ray station (CRS)". 
These stations are connected through the Internet. 
They exchange data in (quasi) real time. 
This makes the system suitable for displaying the arriving cosmic rays
over a wide area at the museums and schools.
The network can be extended by simply adding more stations. 

Detecting time-correlated cosmic rays are usually done by 
observing the signs of the event rate enhancements or observing the chaotic
behavior in the arrival time series.
Both phenomena, however, could also happen accidentally. 
If the observation is done at one location, it is difficult
to distinguish the real ones from fakes. To reject these
fakes, simultaneous measurements at multiple locations are crucial.

\section{Detector}
\label{sec:Detector}

\begin{figure}
\input epsf
\begin{center}
\leavevmode
\epsfxsize=8.0cm
\epsfbox{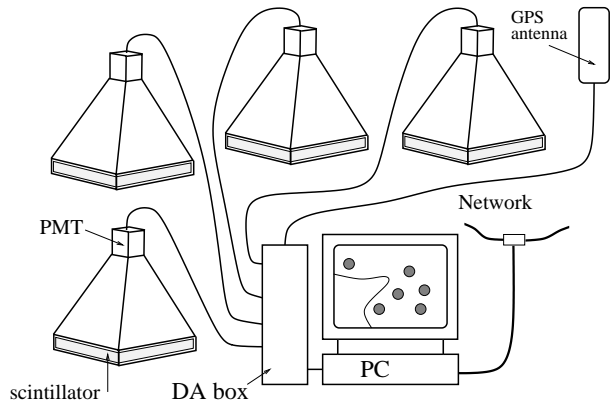}
\caption{Schematics of one Cosmic Ray Station (CRS).
}
\label{fig:one-station}
\end{center}
\end{figure}

\begin{figure}
\input epsf
\begin{center}
\leavevmode
\epsfxsize=12.0cm
\epsfbox{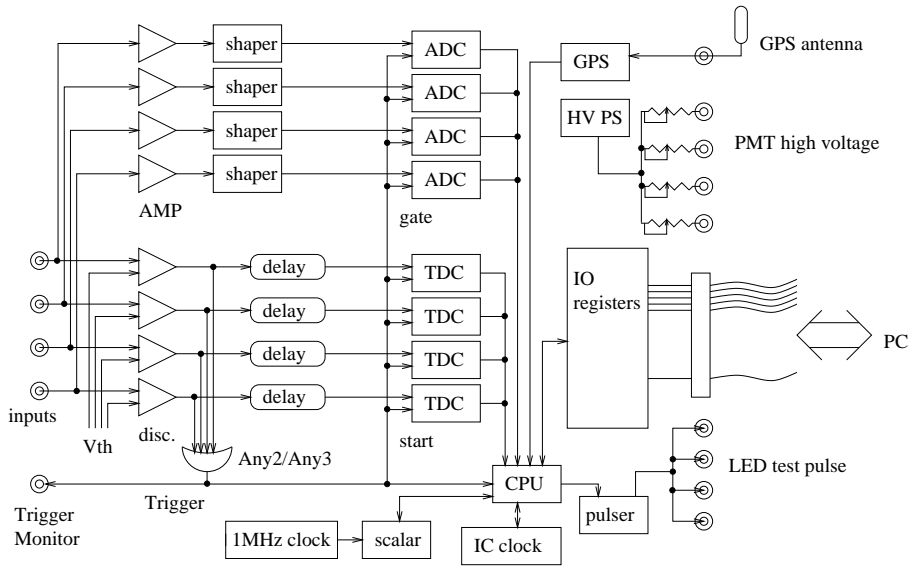}
\caption{Block diagram of the circuit in a DA box.
}
\label{fig:DA-box}
\end{center}
\end{figure}

One station consists of four scintillation counters, a data acquisition 
box (DA box) that has all the electronics for the station, 
a Windows PC with a network interface, and a Global Positioning System 
(GPS) antenna, as shown in Fig. \ref{fig:one-station}. 
Each scintillation counter is comprised
of a pyramid-shaped vessel containing $70 \times 70 \times 4$ cm$^{3}$ 
plastic scintillator at the bottom and a Hamamatsu H6410 2" Photo-Multiplier
Tube (PMT) at the top facing down to the scintillator. 
Each counter is equipped with a Light Emitting Diode (LED) 
for calibration by test pulses. 
The vessel is covered by a soft white polyvinyl chloride sheet for 
protecting the detector from rain.
The DA box has 4 channels of ADCs and TDCs, each having 12-bit resolution, 
to measure the pulse height and timing of the signals 
from the four scintillation counters, as shown in Fig. \ref{fig:DA-box}. 
Triggers are made of coincidences of hits (100 ns width) in either  
two or three out of four counters. The selection is done 
by a switch on the front panel. 
The box has LED drivers, a GPS receiver, a precision clock IC and 
a 1 MHz oscillator with a counter that is synchronized to the GPS signals. 
The system can measure the trigger time with an accuracy of 1 $\mu$s.
The box also has a high voltage power supply for the PMTs. 
The control circuit is made with an Altera Field Programmable Gate Array 
(FPGA) and it is externally controlled by a PC via the IO registers. 

\begin{figure}
\input epsf
\begin{center}
\leavevmode
\epsfxsize=12.0cm
\epsfbox{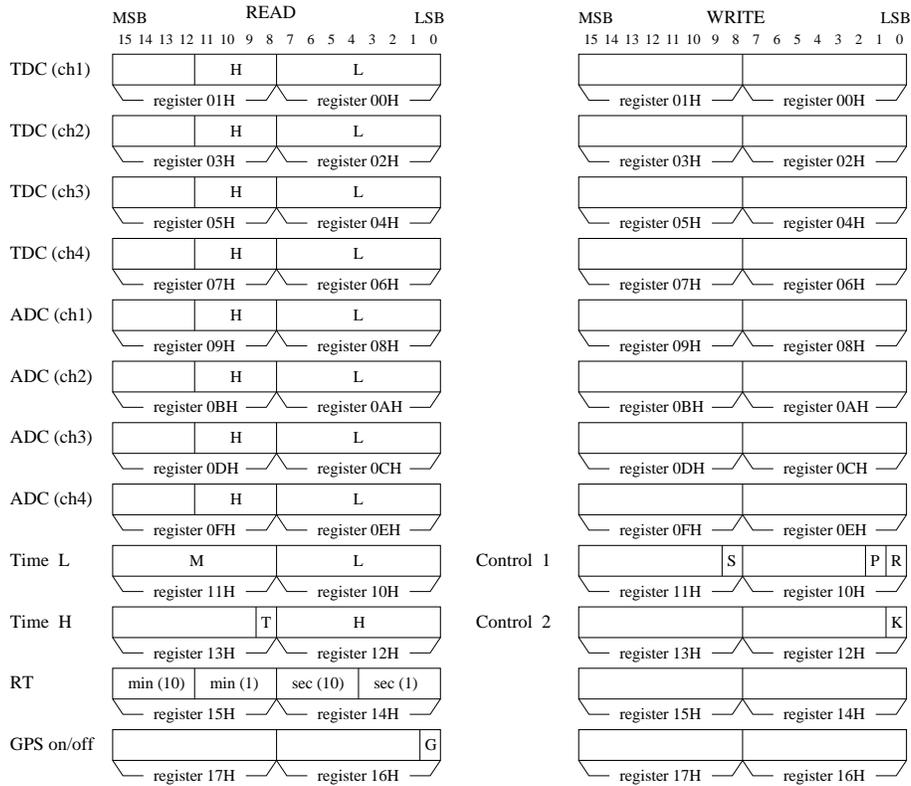}
\caption{Functions and data assignments of the IO registers. 
T: trigger (yes/no)=(1/0), S: run (start/stop)=(1/0), 
P: generate a test pulse (1), R: reset the system (1), 
K: set the clock IC (1).
}
\label{fig:registers}
\end{center}
\end{figure}

The IO registers are used to communicate with the PC through 
a commercially available 16-bit IO board connected to the PCI bus.
The assignments of the functions and data to the IO registers are shown 
in Fig. \ref{fig:registers}. In the READ mode, a trigger (T) 
sets the bit 8 in the ``Time L" register. The trigger time is stored
in the ``RT" register (clock IC) and in the ``Time L" and ``Time H" 
registers (1 $\mu$s counter). 
Four 4-bit data in the ``RT" register, starting from the LSB, show 
the first digit (10$^{0}$) of the seconds, the second 
digit (10$^{1}$) of the seconds, 
the first digit (10$^{0}$) of the minutes, and the second digit (10$^{1}$) of 
the minutes.
The ``Time L" register (L, M) and the low order byte (H) in the 
``Time H" register have the GPS time in $\mu$s, measured by the 1 MHz
oscillator. In the WRITE mode, one can start or stop a run by
setting or resetting bit 8 in the ``Control 1" register.
By setting the bit 1 in the same register, a test pulse is generated.
Bit 0 in the same register is used to reset the system. 
The clock IC is forced to synchronize to the GPS by setting the bit 0
in the ``Control 2" register. 

\begin{figure}
\input epsf
\begin{center}
\leavevmode
\epsfxsize=8.0cm
\epsfclipon
\epsfbox{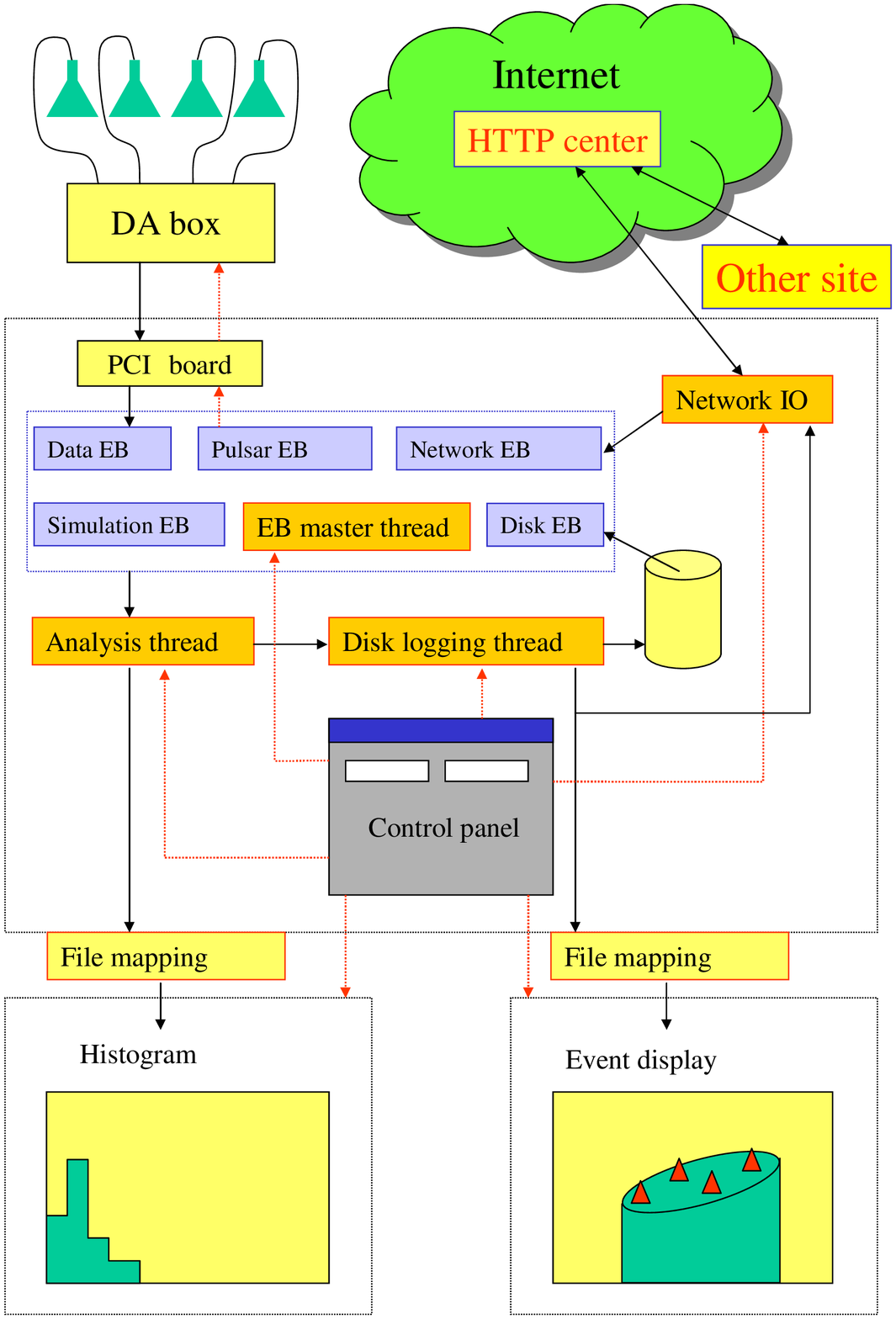}
\caption{Block diagram of the online software. The arrows with solid lines
show the data flows. The command (message) flows are shown by the arrows with 
broken lines.
}
\label{fig:online-soft}
\end{center}
\end{figure}

\begin{figure}
\input epsf
\leavevmode
\epsfxsize=6.8cm
\epsfbox{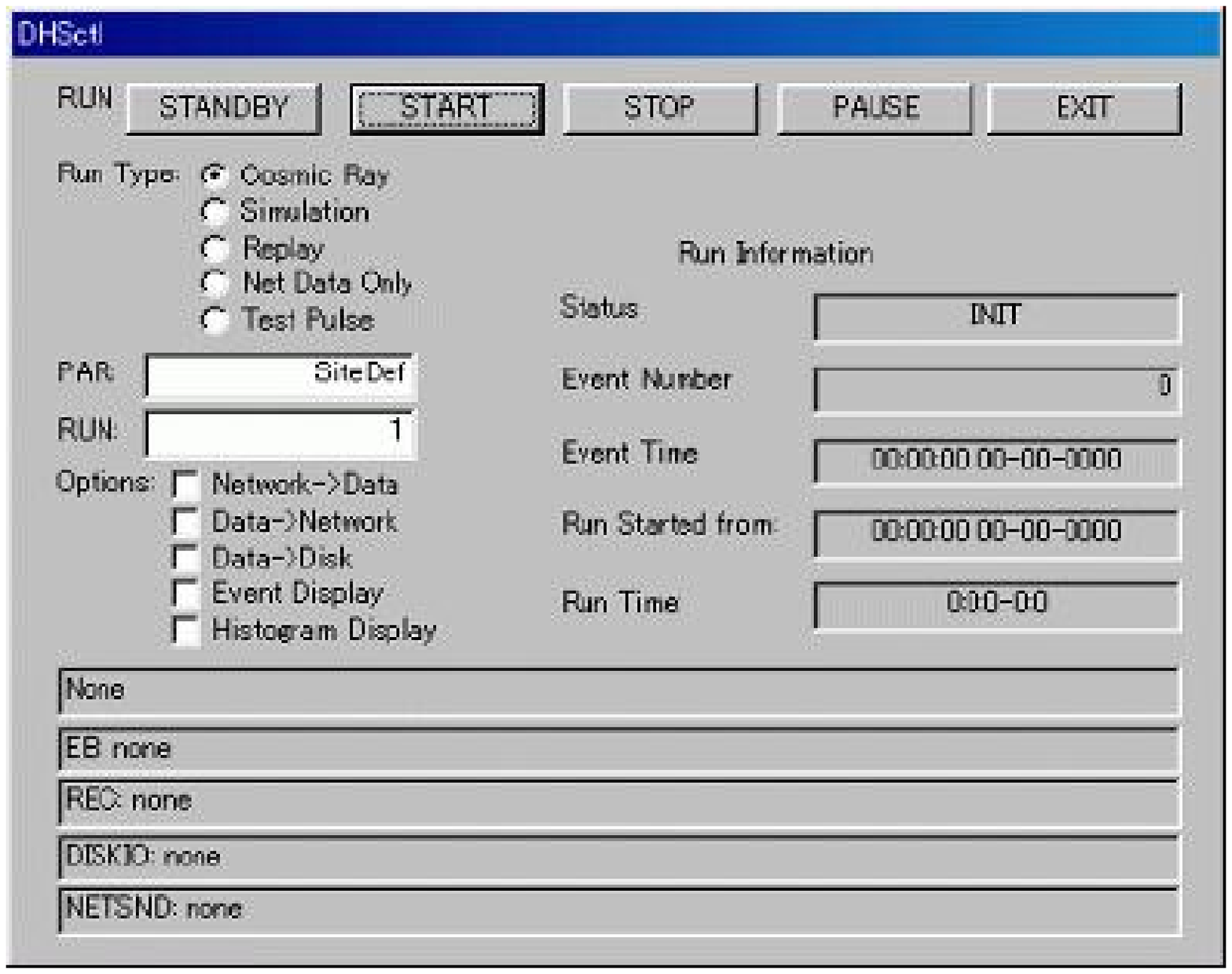}
\epsfxsize=6.8cm
\epsfbox{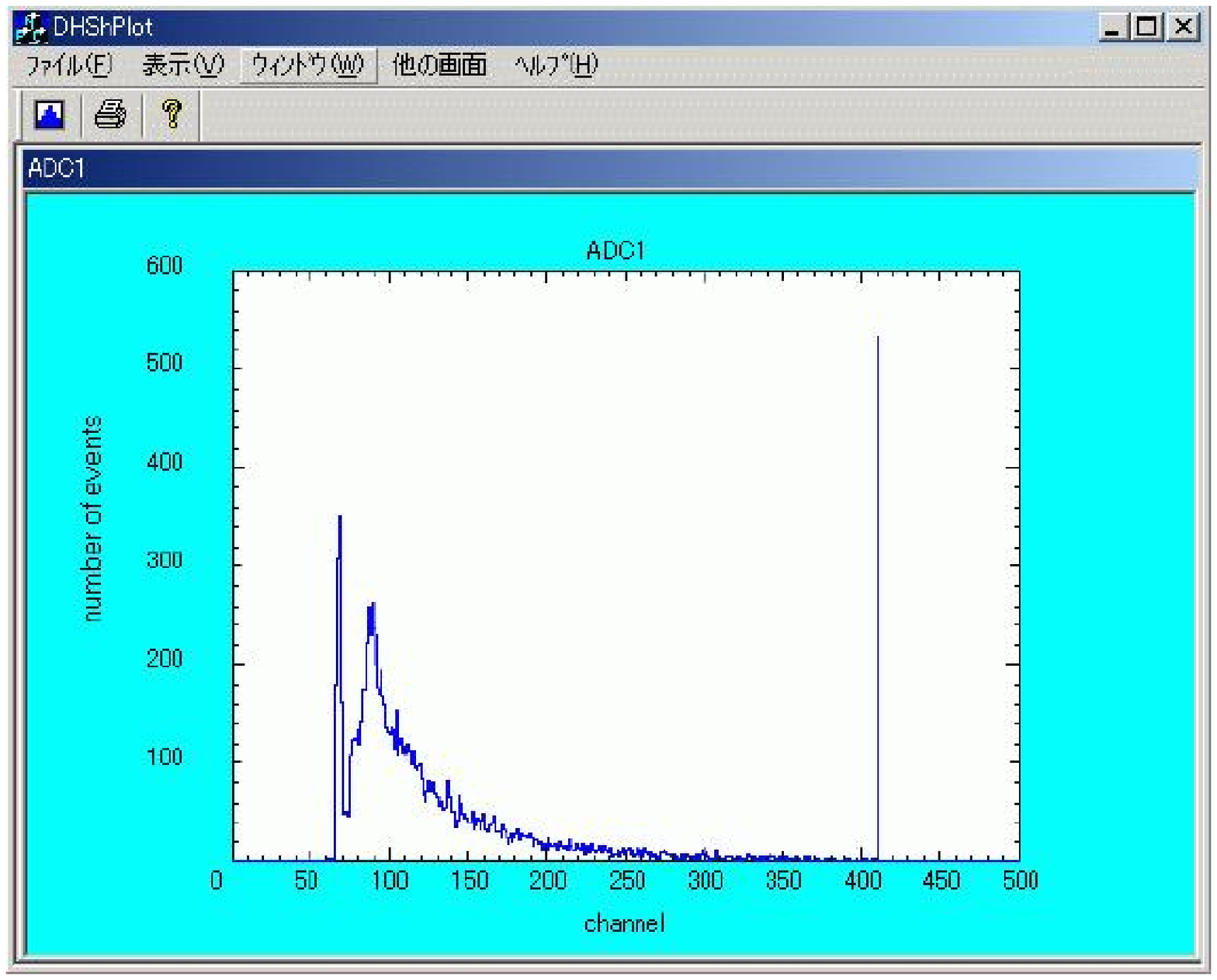}
\\
\epsfxsize=6.8cm
\epsfbox{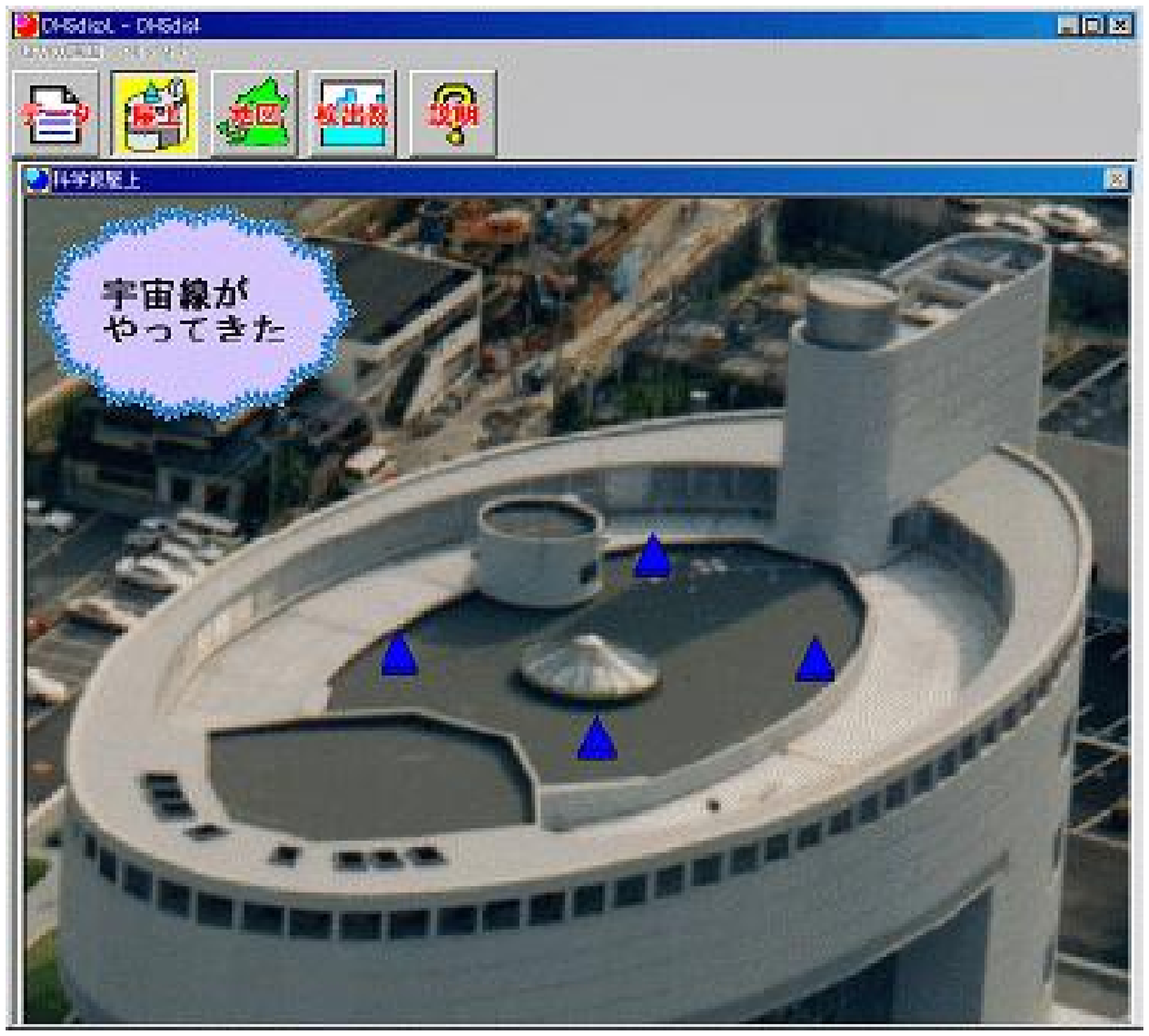}
\epsfxsize=6.8cm
\epsfbox{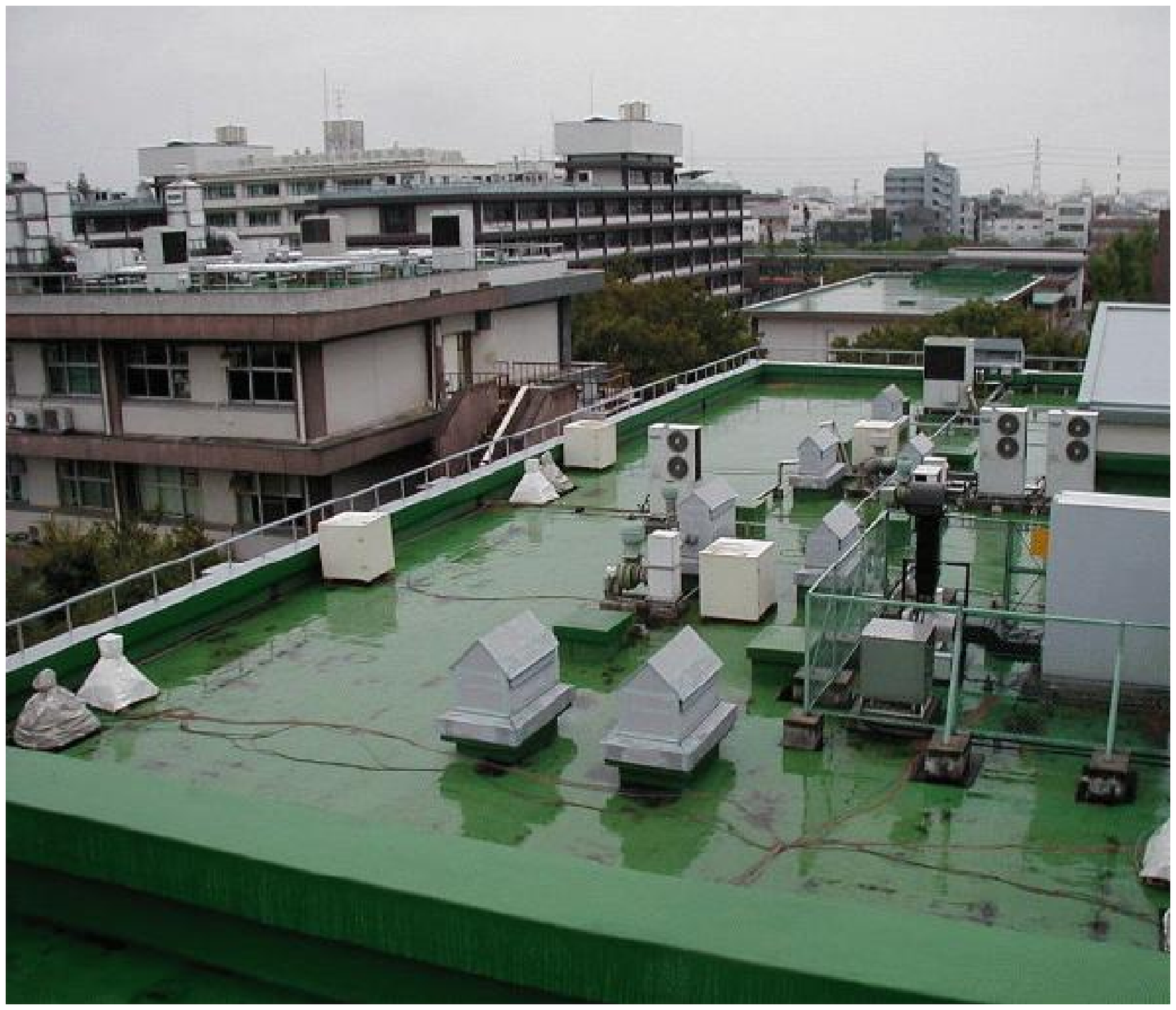}
\begin{center}
\caption{Examples of displays and a detector array. Control panel (top left), 
histogram (top right) and event display (bottom left). 
In the event display, four scintillation counters are shown by small
triangles on the roof. When triggers are made, these triangles change
color for a few seconds with a phrase in Japanese at the top left corner
of the window.
Scintillation counters (ones with brighter white covers) on the roof 
of the Faculty of Science building at the Osaka City 
University (bottom right).
}
\label{fig:DHSctl}
\end{center}
\end{figure}

The online software consists of three processes: a data acquisition process,
an event display process, and a histogram display process, as shown in 
Fig. \ref{fig:online-soft}. They are written in Microsoft 
Visual C++ 6.0 (VC++6.0) and intended to run on a Windows 2000 PC.
The data acquisition process is a multi-thread system. 
It is controlled through a control panel made of a dialog box of 
VC++6.0 (Fig. \ref{fig:DHSctl}). 
The histograms and event displays are multi-window type 
applications of the document-view scheme. 

When the data acquisition process is evoked, it reads a site 
configuration file, which 
is specific to the station. It includes the site ID, group ID,
the site location (latitude, longitude, altitude), scintillator
positions in the local coordinate system, TDC gains, TDC T0s, 
ADC gains, ADC pedestals, and the HTTP\_center's IP address to 
connect. The HTTP\_center is a server process to receive and distribute the
data to and from each station.

Data are handled as C++ objects, which are common 
among the data acquisition program running in a Windows PC and 
the analysis program running in a Linux PC. 
The data format is listed in Table \ref{tab:data-format}.
At transferring data from one thread to
another one in the program, the data objects are queued at each 
receiving thread with a specified depth (typically set to 10). 
This enables us to detect burst-type cosmic ray events. In the event display, 
the depth of the data queue is set to one since we do not have
to display all the events.

\begin{table}
\caption{Data format.}
\label{tab:data-format}
\begin{center}
\hspace*{\baselineskip}
\begin{scriptsize}
\begin{tabular}{|p{2.0cm}|p{2.0cm}|p{2.0cm}|p{2.0cm}|}
\hline
\multicolumn{4}{|c|}{m\_Length} \\
\hline
\multicolumn{4}{|c|}{m\_ID} \\
\hline
\multicolumn{1}{|c|}{m\_GPS\_Stat}
& \multicolumn{1}{|c|}{m\_groupID}
& \multicolumn{2}{|c|}{m\_runN} \\
\hline
\multicolumn{4}{|c|}{m\_eventN} \\
\hline
m\_runType & m\_eventType & m\_Sec & m\_Min \\
\hline
m\_Hour & m\_Day & m\_Month & m\_Year \\
\hline
\multicolumn{4}{|c|}{m\_Latitude} \\
\hline
\multicolumn{4}{|c|}{m\_Longitude} \\
\hline
\multicolumn{4}{|c|}{m\_Altitude} \\
\hline
\multicolumn{4}{|c|}{m\_SciLoc[0].x()} \\
\hline
\multicolumn{4}{|c|}{m\_SciLoc[0].y()} \\
\hline
\multicolumn{4}{|c|}{m\_SciLoc[0].z()} \\
\hline
\multicolumn{4}{|c|}{.} \\
\multicolumn{4}{|c|}{:} \\
\hline
\multicolumn{4}{|c|}{m\_SciLoc[3].x()} \\
\hline
\multicolumn{4}{|c|}{m\_SciLoc[3].y()} \\
\hline
\multicolumn{4}{|c|}{m\_SciLoc[3].z()} \\
\hline
\multicolumn{2}{|c|}{m\_ADCpedestal[0]}
& \multicolumn{2}{|c|}{m\_ADCpedestal[1]} \\
\hline
\multicolumn{2}{|c|}{m\_ADCpedestal[2]}
& \multicolumn{2}{|c|}{m\_ADCpedestal[3]} \\
\hline
\multicolumn{2}{|c|}{m\_TDCtzero[0]}
& \multicolumn{2}{|c|}{m\_TDCtzero[1]} \\
\hline
\multicolumn{2}{|c|}{m\_TDCtzero[2]}
& \multicolumn{2}{|c|}{m\_TDCtzero[3]} \\
\hline
\multicolumn{2}{|c|}{m\_ADC[0]}
& \multicolumn{2}{|c|}{m\_ADC[1]} \\
\hline
\multicolumn{2}{|c|}{m\_ADC[2]}
& \multicolumn{2}{|c|}{m\_ADC[3]} \\
\hline
\multicolumn{2}{|c|}{m\_TDC[0]}
& \multicolumn{2}{|c|}{m\_TDC[1]} \\
\hline
\multicolumn{2}{|c|}{m\_TDC[2]}
& \multicolumn{2}{|c|}{m\_TDC[3]} \\
\hline
\multicolumn{4}{|c|}{m\_ADCgain[0]} \\
\hline
\multicolumn{4}{|c|}{m\_ADCgain[1]} \\
\hline
\multicolumn{4}{|c|}{m\_ADCgain[2]} \\
\hline
\multicolumn{4}{|c|}{m\_TDCgain[3]} \\
\hline
\multicolumn{4}{|c|}{m\_TDCgain[0]} \\
\hline
\multicolumn{4}{|c|}{m\_TDCgain[1]} \\
\hline
\multicolumn{4}{|c|}{m\_TDCgain[2]} \\
\hline
\multicolumn{4}{|c|}{m\_TDCgain[3]} \\
\hline
\multicolumn{4}{|c|}{m\_uSecData} \\
\hline
\multicolumn{2}{|c|}{m\_SecData}
& \multicolumn{2}{|c|}{m\_MinData} \\
\hline
\multicolumn{4}{|c|}{m\_ASzenith} \\
\hline
\multicolumn{4}{|c|}{m\_ASazimuth} \\
\hline
%%\multicolumn{4}{|c|}{m\_ASsize} \\
%%\hline
%%\multicolumn{4}{|c|}{m\_AScoreX} \\
%%\hline
%%\multicolumn{4}{|c|}{m\_AScoreY} \\
%%\hline
\multicolumn{4}{|c|}{m\_ASchi2} \\
\hline
\multicolumn{4}{|c|}{Delineating} \\
\hline
\end{tabular}
\end{scriptsize}
\end{center}
\end{table}

Data exchange between the stations is done through the Internet 
via a server (the HTTP\_center) using HTTP as the protocol. 
The reason for using HTTP is to pass the firewall of the site. 
It can transfer the data through the HTTP proxies to and from the center. 
To send data from a station to the center, the POST command of 
the protocol is issued by the network IO thread in the DA process.
In the center, the received data are stored in a queue.
To receive the data from the center, a polling scheme is used.
The station periodically issues a GET command (currently two seconds in the 
Windows PCs). 
To keep track of the last sent event number to the station, 
the center has a list of entries that have the last event positions in 
the queue for all stations. 
By receiving the GET command, the center looks at this number 
for the station. It then sends all the data stored in the queue 
after that number.
By this scheme, we distribute the data to all stations 
in the ``quasi-real" time. In addition, since the link between the
station and the center is a one-time connection, each station can
start or stop at any time without disturbing the entire observation 
network.

Analysis programs use the same protocol to receive the data 
from the center. It runs in a Linux PC and 
is written with gnu C++ (g++). 
We use an analysis program frame to store (or read) 
the data of all stations in the disk.

\section{Operation and Performance}
\label{sec:performance}

The system has been operated using the stations at the Osaka City University 
and Osaka Science Museum since 2000. At both sites, the scintillation
counters were placed on the roof. They are positioned at the corners of
an approximately 10 $\times$ 10 m$^{2}$ square, as shown 
in Fig. \ref{fig:DHSctl}.

For data taking, the operating voltages of the PMTs are set to 
approximately 1.6 kV and the discriminator levels are set to 25 mV. 
The trigger condition is selected as three-out-of-four.
The trigger rates in both locations are approximately 3 per minute. 

At the museum, only the event displays are shown to the visitors
with the descriptions of the system as well as the general 
cosmic-ray information. The visitors can select the windows using 
the touch panel.

Two more stations were added in 2002 at the Izumiootsu High School and 
Kinki University Technical College. The geometric parameters for
the four sites are listed in Table \ref{tab:site}.

\begin{table}
\caption{Site locations.}
\label{tab:site}
\begin{center}
\begin{tabular}{|l|l|l|l|} \hline
station name & \multicolumn{2}{|c|}{location} & altitude \\ \cline{2-3}
             & latitude & longitude & \\ \hline
Osaka City Univ.	& 34.35'20" & 135.30'20" & sea level \\
Osaka Science Museum	& 34.41'10" & 135.29'30" & sea level \\
Izumiootsu High School	& 34.29'20" & 135.28'00" & sea level \\
Kinki Univ. Technical College
			& 33.40'40" & 136.02'50" & sea level \\ \hline
\end{tabular}
\end{center}
\end{table}

\begin{figure}
\input epsf
\begin{center}
\leavevmode
\epsfxsize=5.0cm
\epsfbox{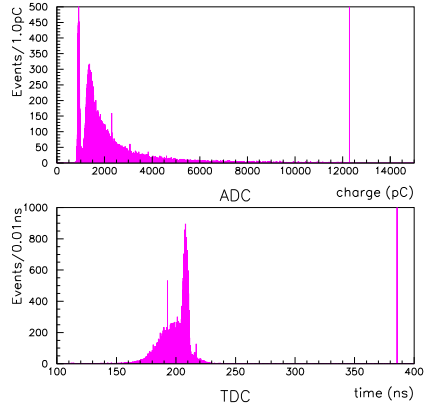}
\caption{ADC (top) and TDC (bottom) of scintillator signals.
}
\label{fig:ADC_TDC}
\end{center}
\end{figure}

\begin{figure}
\input epsf
\begin{center}
\leavevmode
\epsfxsize=5.0cm
\epsfbox{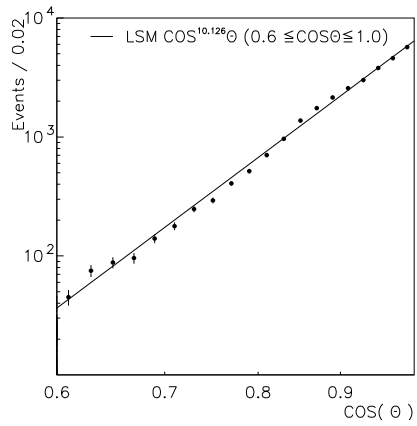}
\caption{Zenith angles.
}
\label{fig:zenith}
\end{center}
\end{figure}

To adjust the operating voltage and to monitor the noise, the ADC data 
(Fig \ref{fig:ADC_TDC}) of each scintillation counter at each site 
are monitored via the network at the Osaka City University. 
TDC data are used to
obtain the arrival directions of the air showers assuming the shower fronts
are flat. The sharp bump in the TDC distribution at 210 ns in 
Fig. \ref{fig:ADC_TDC} shows the events that the signal of this channel 
was the third in arrival time among the four channels; each corresponds 
to one scintillation counter. Since a trigger is generated 
by the signals in any three channels, the signal timing of this 
channel determined the trigger timing, thus giving the constant TDC count. 
The broad distribution is made of the signals arriving in the other 
time ordering (i.e. ``non-third" in the arrival time).
The front planes of the showers are obtained by the least squares
method. From the residual of the fit, the angle resolution was obtained
as 8$^\circ$. 
The measured zenith angles, $\theta$'s, are shown in Fig. \ref{fig:zenith}.
The distribution can be fit by $\cos^{n}\theta$ with $n=10.1$ for 
$0.6 \le \cos\theta \le 1.0$. The zenith angles $\cos\theta < 0.6$
has a flatter distribution. Therefore, we only accepted the events 
with $\cos\theta \ge 0.6$ for the analysis.
After this cut, the background are estimated to be less than 
a few percent.

\begin{figure}
\input epsf
\begin{center}
\leavevmode
\epsfxsize=5.0cm
\epsfbox{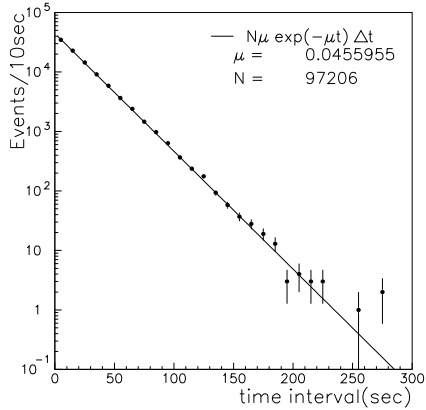}
\caption{Arrival time intervals at Osaka City University.
}
\label{fig:arrival}
\end{center}
\end{figure}

\begin{figure}
\input epsf
\begin{center}
\leavevmode
\epsfxsize=5.0cm
\epsfbox{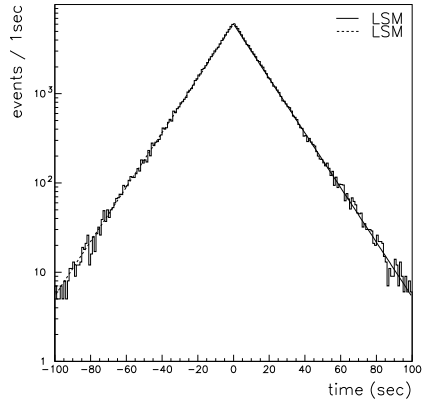}
\caption{Time correlation between the events at the Osaka City University and
Osaka Science Museum.
}
\label{fig:time-correlation}
\end{center}
\end{figure}

The arrival time distribution is shown in Fig. \ref{fig:arrival}
using the events observed at the Osaka City University. 
The exponential behavior shows that the events are randomly observed 
with a rate of 2.7 per minute.
Figure \ref{fig:time-correlation} shows the arrival time interval of the
events at the Osaka Science Museum with respect to the ones at the Osaka 
City University. The used data were collected at the Osaka City University
via the Internet for a period of one month.
The negative or positive time in the distribution show the cases 
when the events at the museum are earlier (negative) or later (positive) 
than the ones at the university.
This indicates that the events at both locations were observed by
the network data-taking with no biasing. 

\section{Summary}
\label{sec:summary}

We have developed a cosmic ray detection system suitable for detecting 
air showers arriving over a wide area with correlated time. The
system consists of detector stations connected by the Internet. Each
station is comprised of four scintillation counters with an electronics box 
and a PC, which can reconstruct the local air showers. For data communication,
HTTP is used as the protocol to pass the firewalls. 
Each station polls the HTTP server at the Osaka City University, 
which collects and distributes the data to each site. 

The system is currently operating with four sites. It can be expanded by
simply adding stations. A typical trigger rate at each station is 3 
per minute. 
The measured arrival direction resolution is 8$^{\circ}$. 
The measured arrival time of cosmic rays are consistent with random, 
indicating that the system has no biasing.
We plan to expand the system and cover a larger detection area in order 
to search for time-correlated events whose frequency is expected to 
be less than a few per year.

\noindent{\em Acknowledgments.}
\smallskip \\
\hspace*{12pt}We thank Prof. T.~Kitamura and W.~Unno of Kinki 
University for encouraging us to build this system. We also thank
Prof. H.~Fujii of the National Laboratory for High Energy Physics (KEK)
for the useful discussions in developing the software. 
This project is partially funded by a Grant-in-Aid for Scientific 
Research on Priority Areas
by the Japanese Ministry of Education, Culture, Sports, Science and 
Technology (MEXT).

\end{document}